\documentclass[aps,prd,twocolumn,nofootinbib]{revtex4}
\usepackage{graphicx}
\newcommand{\beq}{\begin{equation}}
\newcommand{\eeq}{\end{equation}}
\newcommand{\bqa}{\begin{eqnarray}}
\newcommand{\eqa}{\end{eqnarray}}
\begin{document}

\title{Estimation of Semileptonic Decays of $B_c$ Meson to S-wave Charmonia with NRQCD}

\author{Cong-Feng Qiao$^{1,2}$}
\email[email:]{qiaocf@gucas.ac.cn}

\author{Rui-Lin Zhu$^{1}$}
\email[email:]{zhuruilin09@mails.gucas.ac.cn}

\address{$^{1}$Department of Physics,
University of the Chinese Academy of Sciences,
\\YuQuan Road 19A, Beijing 100049, China\\
$^{2}$Kavli Institute for Theoretical Physics China, the Chinese
Academy of Sciences, Beijing 100190, China}


\begin{abstract}

We study the semileptonic differential decay rates of $B_c$ meson to
S-wave charmonia, $\eta_c$ and $J/\Psi$, at the next-to-leading
order accuracy in the framework of NRQCD. In the heavy quark limit,
$m_b \to \infty$, we obtain analytically the asymptotic expression
for the ratio of NLO form factor to LO form factor. Numerical
results show that the convergence of the ratio is perfect. At the
maximum recoil region, we analyze the differential decay rates in
detail with various input parameters and polarizations of $J/\psi$,
which can now be checked in the LHCb experiment. Phenomenologically,
the form factors are extrapolated to the minimal recoil region, and
then the $B_c$ to charmonium semileptonic decay rates are estimated.

\begin{description}
\item[PACS numbers] 12.38.Bx, 12.39.St, 13.20.-v
\end{description}

\end{abstract}

\maketitle

\section{Introduction}
Hadron collider provides a large amount of data on $B_c$ events.
Wherein the most easily identified decay modes to tag the $B_c$ are:
fully reconstructed channel $B_{c}\to J/\Psi \pi$ and semileptonic
decay channel $B_{c}\to J/\Psi \ell \nu(\ell =e,\mu)$. The CDF
Collaboration made the first observation of the $B_c$ meson by its
semileptonic decay at the Tevatron fourteen years ago
\cite{Bc:1998}. Latter, the D0 Collaboration performed the same
analysis in a sample of $210~\mathrm{pb}^{-1}$ of the Run II data
\cite{Bc:2005}. The cross section of $B_c$ production at the Large
Hadron Collider(LHC) is larger than that at the Tevatron by roughly
an order of magnitude, which reaches $49.8~\mathrm{nb}$ at the
center-of-mass energy $\sqrt{s}=14~\mathrm{TeV}$
\cite{HQP,BCevent2}. This makes the experimental study of the
differential branching fraction of $B_c$ meson semileptonic decays
to charmonium feasible. We can also obtain the information of the
Cabibbo-Kobayashi-Maskawa(CKM) matrix element in $B_c$ decays,
especially $V_{cb}$ which is not well determined.

Recently, the BABAR collaboration measured the partial branching
fraction  $\Delta B/\Delta q^2$ in bins of the momentum-transfer
squared, with $6~q^2$ bins for $B^0\to\pi^{-}\ell^{+}\nu$ and
$3~q^2$ bins for $B^0\to\rho^{-}\ell^{+}\nu$ \cite{rho:2011}. They
found that the partial branching fraction of
$B^0\to\pi^{-}\ell^{+}\nu$ decreases as $q^2$ increases, while for
$B^0\to\rho^{-}\ell^{+}\nu$ process, the partial branching fraction
increases first and then decreases as $q^2$ increases. Actually, we
know that all of the five form factors in above two decay channels
at the maximum recoil region increases with $q^2$, at the
next-to-leading order(NLO) accuracy according to the light cone sum
rules calculation \cite{piform,rhoform}. The decrease of
$B^0\to\pi^{-}\ell^{+}\nu$ is caused by the phrase space, which
counteracts the enhancement from form factors. In this work, we try
to make out whether this happens or not in $B_c$ semileptonic decays
to charmonia.

There exist several approaches in the calculation of $B_c$ meson
semileptonic decays to charmonium. Some of them are: the light cone
QCD sum rules
\cite{Kiselev:2000,Kiselev:2004,Huang:2007,Huang:2008}, the
relativistic quark model \cite{Ivanov:2001, Ebert:2003}, the
instantaneous non-relativistic approach to the Bethe-Salpeter
equation \cite{Zhang:1994}, the non-relativistic constituent quark
model\cite{cqm}, the covariant light front model \cite{Lu:2009}, and
the QCD potential model \cite{QCDP}.

Consider that the $B_c$ meson is constituted by two heavy quarks
with different flavors, which masses are much larger than the
$\Lambda_{QCD}$, analogous to the situation of heavy quarkonium, the
system turns out to be non-relativistic. Hence the relative velocity
of heavy quarks within the $B_c$ meson is small, i.e. $\upsilon\ll
1$, though bigger than the velocities of quarks in charmonium and
bottomonium systems, and the non-relativistic QCD(NRQCD) formalism
is applicable to the study of $B_c$ meson semileptonic decays to
charmonia. In the NRQCD framework, the matrix elements of the
concerned processes can be factorized as
\begin{eqnarray}
\langle J/\psi(\eta_c)\ell \nu|\overline{c} \Gamma_\mu
b\overline{\ell} \Gamma^\mu
\nu|B_c\rangle\simeq\sum_{n=0}\psi(0)_{B_c}\psi(0)_{J/\psi(\eta_c)}T^n\;
.
\end{eqnarray}
Here, $\Gamma^\mu=\gamma^{\mu}(1-\gamma_{5})$, the nonperturbative
parameters $\psi(0)_{\bar{B}_c}$ and $\psi(0)_{J/\psi(\eta_c)}$ are
the Schr\"{o}dinger wave functions at the origin for $b\bar{c}$ and
$c\bar{c}$ systems, respectively. $T^n$ are hard scattering kernels
which can be calculated perturbatively.

The paper is organized as follows: In section~\ref{sec-form1}  we
present the definition for relevant form factors and work out the
expressions of form factors in the NRQCD framework. In
section~\ref{sec-form} the dependence of the NLO semileptonic
differential decay rates on $q^2$ is  obtained.  In
section~\ref{sec-decaywidth} we calculate the decay width and study
the theoretical uncertainty, and analyze the result in detail of the
maximum recoil region. The last section is remained for conclusions.

\section{Form factors\label{sec-form1}}

The  $B_{c}\to J/\psi (\eta_{c})$ transition form factors, $f_{+}$,
$f_{0}$, $V$, $A_{0}$, $A_{1}$, and $A_{2}$  are normally defined as
follows \cite{Wirbel}
\begin{eqnarray}
\langle \eta_{c}(p)\vert \bar c \gamma^{\mu}b\vert B_{c}(P)\rangle
&=&f_{+}(q^{2})(P^{\mu}+p^{\mu}-\frac{m_{B_{c}}^{2}-
m_{\eta_{c}}^{2}}{q^{2}}q^{\mu})\nonumber\\ &&+f_{0}(q^{2})
\frac{m_{B_{c}}^{2}-m_{\eta_{c}}^{2}}{q^{2}}q^{\mu}\,,
\end{eqnarray}
\begin{eqnarray}
 && \langle J/\psi(p,\varepsilon^{*})\vert \bar c \gamma^{\mu}b\vert
B_{c}(P)\rangle =\frac{2 i
V(q^{2})}{m_{B_{c}}+m_{J/\psi}}\epsilon^{\mu\nu\rho\sigma}
\varepsilon_{\nu}^{*}p_{\rho}P_{\sigma}\,,
\nonumber\\
&& \langle J/\psi(p,\varepsilon^{*})\vert \bar c
\gamma^{\mu}\gamma_{5}b\vert B_{c}(P)\rangle =2
m_{J/\psi}A_{0}(q^{2})\frac{\varepsilon^{*}\cdot q}{q^{2}}q^{\mu}
\nonumber\\&&~~~~~-A_{2}(q^{2})\frac{\varepsilon^{*}\cdot
q}{m_{B_{c}}+m_{J/\psi}}(
P^{\mu}+p^{\mu}-\frac{m_{B_{c}}^{2}-m_{J/\psi}^{2}}{q^{2}}q^{\mu})
\nonumber\\&&~~~~~+(m_{B_{c}}+m_{J/\psi})A_{1}(q^{2})
(\varepsilon^{*\mu}-\frac{\varepsilon^{*}\cdot q}{q^{2}} q^{\mu})\,.
\end{eqnarray}
Here we define the momentum transfer $q=P-p$.

It is straightforward to calculate those form factors at the tree
level in the NRQCD. They read
\bqa V^{LO}(q^{2})=\frac{16 \sqrt{2} C_A C_F \pi  (3 z+1) \alpha _s
\psi(0)_{B_c}\psi(0)_{J/\Psi}}{\left((1-z)^2-\frac{q^2}{m_b^2}\right)^2
   \left(\frac{z}{z+1}\right)^{3/2} m_b^3 N_c}\; ,
 \eqa
\bqa A_0^{LO}(q^2)=\frac{16 \sqrt{2} C_A C_F \pi  (z+1)^{5/2} \alpha
_s\psi(0)_{B_c}\psi(0)_{J/\Psi}}{\left((1-z)^2-\frac{q^2}{m_b^2}
\right)^2 z^{3/2} m_b^3 N_c}\; ,\nonumber\\
 \eqa

\begin{widetext}
\bqa A_1^{LO}(q^2)=\frac{16 \sqrt{2} C_A C_F \pi  \sqrt{z+1} \left(4
z^3+5 z^2+6 z-\frac{q^2}{m_b^2} (2 z+1)+1\right) \alpha _s
\psi(0)_{B_c}\psi(0)_{J/\Psi}}{\left((1-z)^2-\frac{q^2}{m_b^2}\right)^2
z^{3/2} (3 z+1) m_b^3 N_c}\; ,
 \eqa
\bqa A_2^{LO}(q^2)=\frac{16 \sqrt{2} C_A C_F \pi  \sqrt{z+1} (3 z+1)
\alpha_s\psi(0)_{B_c}\psi(0)_{J/\Psi}}{\left((1-z)^2-
\frac{q^2}{m_b^2}\right)^2
z^{3/2} m_b^3 N_c}\; ,
 \eqa
\bqa f_+^{LO}(q^{2})=\frac{8 \sqrt{2} C_A C_F \pi  \sqrt{z+1}
\left(-\frac{q^2}{m_b^2}+3 z^2+2 z+3\right) \alpha _s \psi(0)_{B_c}
\psi(0)_{\eta_c}}{\left((1-z)^2-\frac{q^2}{m_b^2}\right)^2 z^{3/2}
m_b^3 N_c}\; , \eqa
\bqa f_0^{LO}(q^{2})=\frac{8 \sqrt{2} C_A C_F \pi  \sqrt{z+1}
\left(9 z^3+9 z^2+11 z-\frac{q^2}{m_b^2} (5 z+3)+3\right) \alpha _s
\psi(0)_{B_c}
\psi(0)_{\eta_c}}{\left((1-z)^2-\frac{q^2}{m_b^2}\right)^2 z^{3/2}
(3 z+1) m_b^3 N_c}\; ,
 \eqa
\end{widetext}
where $z\equiv m_c/m_b$.

There are three typical scales of the process, which possess the
hierarchy of $\Lambda_{QCD}\ll m_c \ll m_b$. Note that in
Ref.~\cite{Bell:2007, Bell:2007-2,Sun:2011}, the form factors of
$B_c$ transition to $\eta_c$ or $J/\Psi$ with alternative
parameterizations have been calculated at the NLO accuracy in the
non-relativistic limit.  We expand the ratios of the NLO form
factors to the leading order(LO) form factors at first order in $z=
m_c/m_b$ expansion in the heavy quark limit $m_b \to \infty$. And
the asymptotic expressions of which are then obtained analytically,
that can be found in the Appendix A.

In the heavy quark limit, the form factors become
 \bqa
V(q^{2})^{LO}_{m_b \rightarrow\infty}=\frac{16 \sqrt{2} C_A C_F \pi
\alpha _s
\psi(0)_{B_c}\psi(0)_{J/\Psi}}{\left(1-\frac{q^2}{m_b^2}\right)^2
   z^{3/2} m_b^3 N_c}\; ,
 \eqa
\bqa A_2(q^2)_{m_b \rightarrow\infty}
=V(q^2)_{m_b \rightarrow\infty}\; ,
 \eqa
 \bqa A_0(q^2)^{LO}_{m_b \rightarrow\infty}
=V(q^2)^{LO}_{m_b \rightarrow\infty}\; ,
 \eqa
\bqa A_1(q^2)_{m_b \rightarrow\infty}
=\left(1-\frac{q^2}{m_b^2}\right)V(q^2)_{m_b
\rightarrow\infty}\; ,
 \eqa
\bqa f_+^{LO}(q^{2})_{m_b \rightarrow\infty}=\frac{
\left(3-\frac{q^2}{m_b^2}\right)
\psi(0)_{\eta_c}}{2\psi(0)_{J/\Psi}}V(q^2)^{LO}_{m_b
\rightarrow\infty}\; , \eqa
 \bqa f_0^{LO}(q^{2})_{m_b
\rightarrow\infty}=\frac{ 3\left(1-\frac{q^2}{m_b^2}\right)
}{\left(3-\frac{q^2}{m_b^2}\right)}f_+^{LO}(q^{2})_{m_b
\rightarrow\infty}\; . \eqa %
At $q^2=0$ point, some form factors turn to be identical, that is:
 \bqa f_0(0)=f_+(0)\;\label{ff1} ,
 \eqa
\bqa V(0)_{m_b \rightarrow\infty}=A_1(0)_{m_b
\rightarrow\infty}=A_2(0)_{m_b \rightarrow\infty}\; ,
 \eqa
which are in consistent with the Heavy Quark Effect Theory (HQET)
\cite{HQET} and the Large Energy Effective Theory (LEET) \cite{LEET}
predictions. Note that the equality (\ref{ff1}) still holds beyond
the heavy limit.

While approaching to the minimal recoil region, the charmonium will
keep still in rest frame of initial particle, meanwhile the
invariant mass of lepton and neutrino pair will turn to its maximum
value. In this case, the gluon exchanged inside the hadrons becomes
soft, which may result in infinity in the evaluation.

To extrapolate the form factors to the minimal recoil region, there
exist several different approaches in the literature. One of them is
the pole mass dependence model developed in
Refs.~\cite{Lu:2009,exBc,narod}, where the form factors are
parametrized as
\begin{equation}\label{pole mass}
    f^\prime(q^2)=\frac{f(0)}{1-q^2/m^2_{\mathrm{pole}}-
    \beta q^4/m^4_{\mathrm{pole}}}\; .
\end{equation}
Here, $\beta$ is free parameter, which is set to be zero in our
calculation as did in Ref.~\cite{exBc}; $m_{\mathrm{pole}}$ denotes
the gluon effective pole mass; $f^\prime(q^2)$ represents any one of
the form factors. In the latter calculation for decay widths, we
will adopt this form. To regulate the infrared divergence induced by
the soft gluon, one asks form factors $f^\prime(q^2)$ satisfying
conditions
\begin{equation}\label{soft}
   f^\prime(q^2)_{q^2\rightarrow 0}=f(0),~~~ f^\prime(q^2)_{q^2
    \rightarrow q^2_{\mathrm{max}}}=\mathrm{constant}\; .
\end{equation}
Here the constant represents the value of form factors at the
minimal recoil point and may be determined through certain model.
For example, we can parameterize the form factors as
\begin{equation}\label{soft}
    f^\prime(q^2)=f(\frac{q^2}{\sqrt{1+(q^2/q^2_{cut})^2}})e^{-S(q^2)}\;
    ,
\end{equation}
and it satisfies $f^\prime(q^2)\simeq f(q^2)$ in the maximum recoil
region, while becomes finite at the minimal recoil point. Here, the
$S(q^2)$ meets the condition $S(0)=0$ and hence can be further
parameterized as $S(q^2)=c_0 q^2$ with $c_0$ a constant. $q^2_{cut}$
is introduced to regularize the unphysical behavior of form factors
in the minimal recoil region. Note that the parameters $q^2_{cut}$
and $c_0$ should be either determined through phenomenological model
or fitted by experimental data.

\section{ semileptonic differential decay widths\label{sec-form}}

For light leptons $e$ and $\mu$, their masses $m_\ell$ can be
readily neglected, hence the semileptonic differential decay rate of
$B_{c}\to \eta_c \ell \nu$ depending on $q^2$ reads
 \begin{eqnarray}
\label{gammaq2} \frac{d\Gamma}{dq^2}(B_{c}\to \eta_c \ell
\nu)=\frac{G^2_F |V_{cb}|^2}{192 \pi^3 m_{B_c}^3} \lambda(q^2)^{3/2}
   [f_+(q^2)]^2\; .
\end{eqnarray}
Here, $G_F$ is the Fermi constant; $V_{cb}$ is the CKM matrix
element; and $\lambda(q^2)=(m_{B_c}^2+m_{\eta_c}^2 -
q^2)^2-4m_{B_c}^2 m_{\eta_c}^2$. For lepton $\tau$, its mass can not
be ignored in the analysis, in which the form factors $f_0$ and
$f_+$ are both involved in. However, $f_0$ can be measured via
$B_{c}\to \eta_c \tau \nu_\tau$ process, while $f_+$ can be obtained
through $B_{c}\to \eta_c \ell \nu_\ell$ decay.

\begin{figure}
\includegraphics[width=0.40\textwidth]{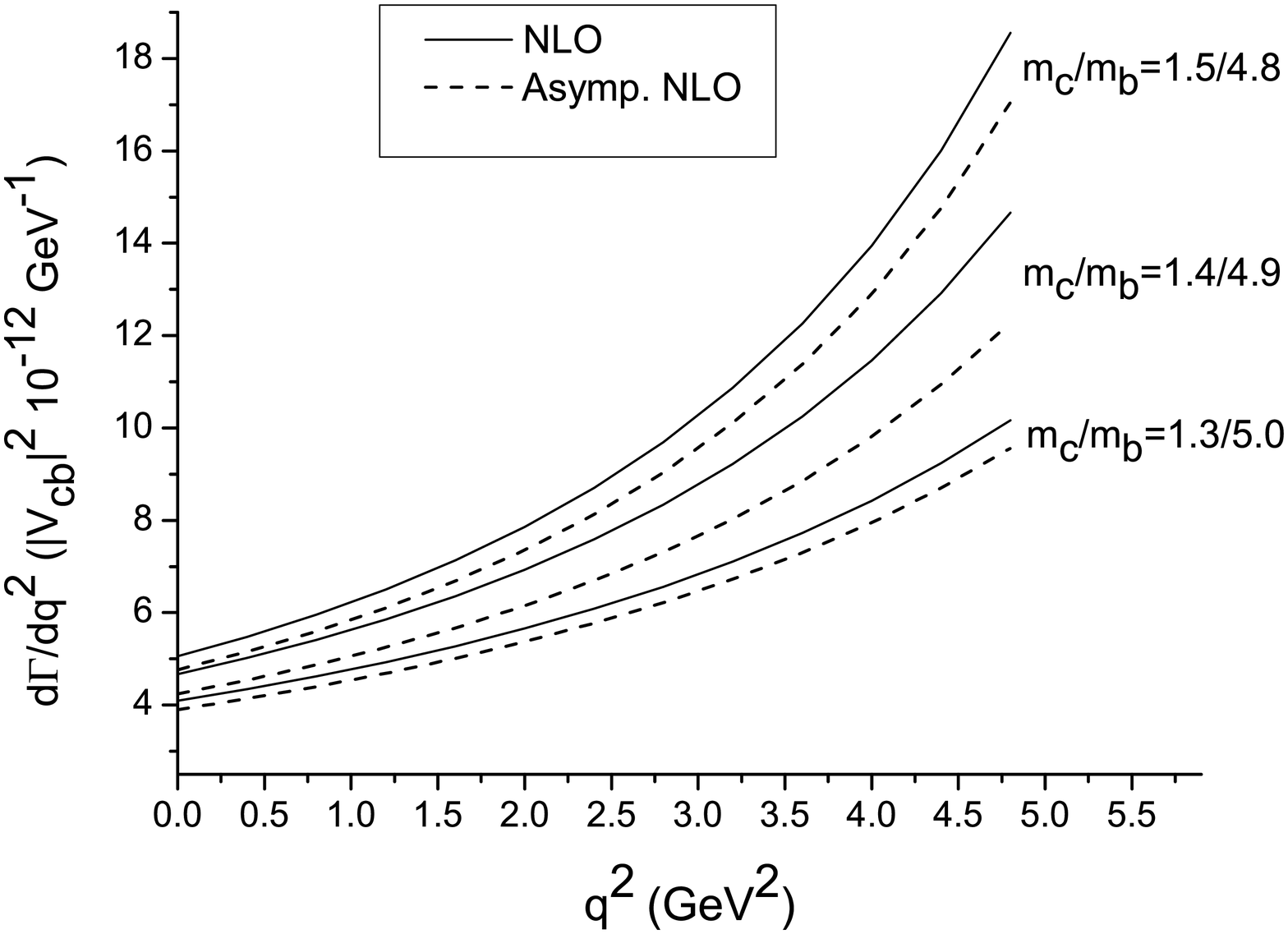}
\caption{NLO differential decay rate for the $B_{c}\to \eta_c
\ell\nu$, for different values of quark mass. The renormalization
scale is chosen to be close to the bottom quark mass, i.e.
$\mu=4.8~\mathrm{GeV}$. In the figure, Asymp.~NLO means expanding
the ratio of NLO form factor to LO one at the first order in $z=
m_c/m_b$ expansion and in the heavy quark limit $m_b \to \infty$.}
\label{Fig:Semilep-etac}
\end{figure}

\begin{figure}
\includegraphics[width=0.40\textwidth]{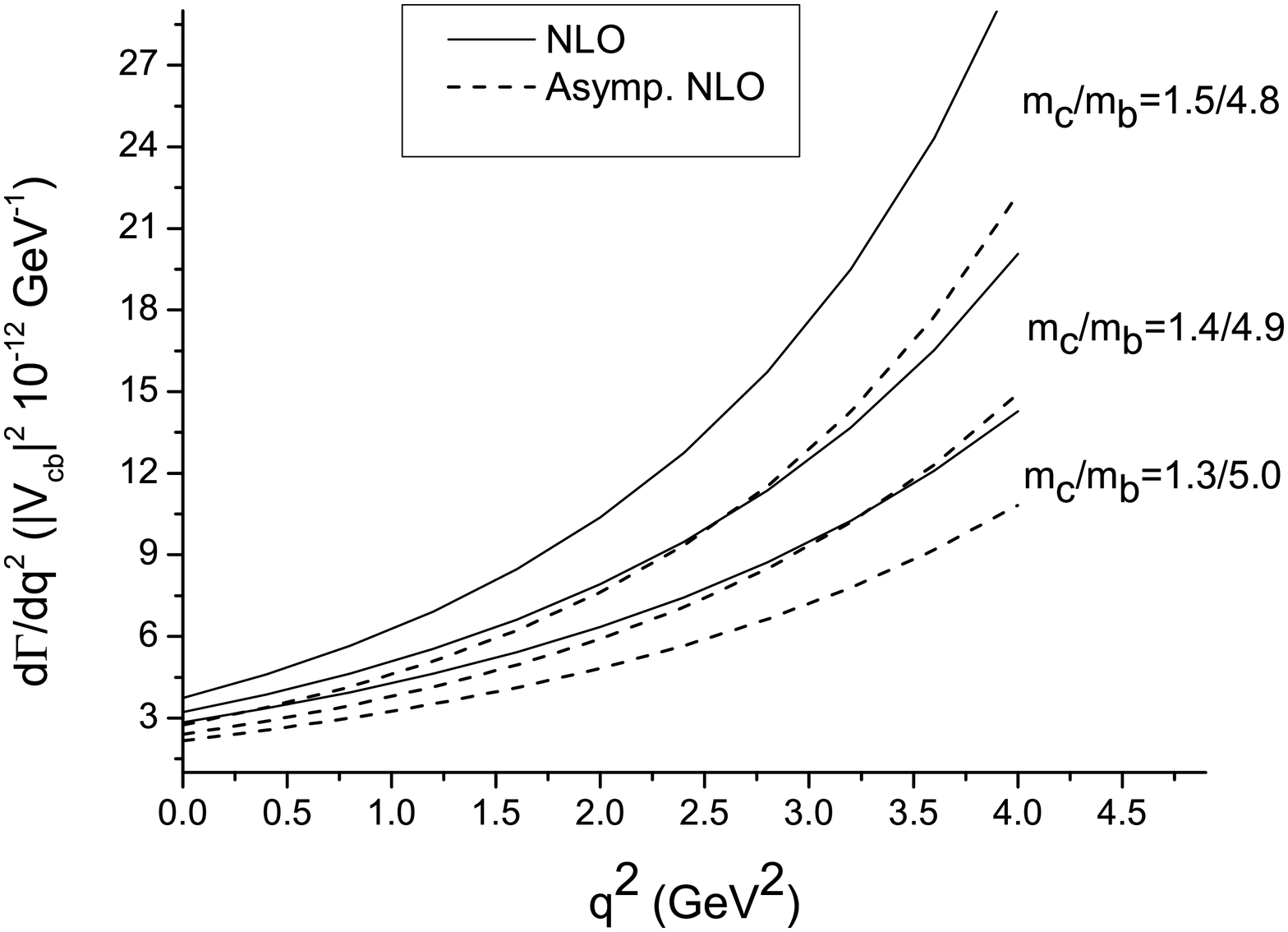}
\caption{NLO differential decay rate for the $B_{c}\to J/\Psi
\ell\nu$, for different values of quark mass. The renormalization
scale is chosen to $\mu=4.8~\mathrm{GeV}$.} \label{Fig:Semilep}
\end{figure}

By virtue of the NLO form factors, we can easily gain the
distribution of NLO differential decay rate on momentum transfer
$q^2$. To check the convergence behavior of the ratio of NLO
differential decay rate to LO one, we select three sets of different
values of $z$ and scale $q^2$, as given in Table \ref{tab:Bc-lep1},
and illustrate parameter dependence in Fig.~\ref{Fig:Semilep-etac}
and \ref{Fig:Semilep}. Here, the Schr\"{o}dinger wave function at
the origin for $J/\Psi$ is determined through its leptonic decay
widths at the NLO level.

For $B_{c}\to \eta_c \ell\nu$ channel, at the maximum recoil point
$q^2=0$, we obtain a value of $4.67^{+0.38}_{-0.58}\times
10^{-12}|V_{cb}|^2 ~\mathrm{GeV}^{-1}$ for (\ref{gammaq2}), which is
larger than the value of $2.05\times 10^{-12}|V_{cb}|^2
~\mathrm{GeV}^{-1}$ obtained in QCD LCSR\cite{Huang:2008} and the
value of $0.65\times 10^{-12}|V_{cb}|^2 ~\mathrm{GeV}^{-1}$ obtained
in nonrelativistic quark model \cite{cqm}. Besides, the results away
from the maximum recoil point tend to disagree with what in
Ref.~\cite{cqm}. In NRQCD calculation, the form factors of $B_c$ to
$\eta_c$ are obviously enhancing with $q^2$ increase, other than
results from light cone sum rules and nonrelativistic quark model,
and the trend is sharpening at NLO, which counteracts the
decrescence due to the factor of phrase space.

\begin{table}[t]
\caption{\label{tab:Bc-lep1} Theoretical parameters for different
sets, with renormalization scale $\mu=4.8~\mathrm{GeV}$, the
lifetime of the $B_c$ $\tau(B_c) =0.453$ ps, and
$G_F=1.16637\times10^{-5}~\mathrm{GeV}^{-2}$ \cite{PDG:2010}, where
$m_b$, $m_c$ and $\Lambda$ are in unit of $\mathrm{GeV}$, while
$|\psi(0)|s$ are in unit of $\mathrm{GeV}^{3/2}$
\cite{Eichten,Qiao:2011}.}
\begin{center}
\begin{tabular}{|c|c|c|c|c|c|c|}
\hline
  & $m_b$  & $m_c$  &
 $\Lambda$  & $|\psi(0)|_{B_c}$ & $|\psi(0)|_{\eta_c}$
  & $|\psi(0)|_{J/\Psi}$ \\
\hline
 set 1 & 4.8 & 1.5   &   &
 &  &    \\

 set 2  & 4.9   & 1.4  &
0.10 & 0.3615 & 0.283  &0.283\\
 set 3  &5.0    & 1.3  &   &
 &  & \\
\hline
\end{tabular}
\end{center}
\end{table}

For the channel of $B_{c}\to J/\Psi \ell\nu$($\ell =e,\mu$), the
decay rates in transverse and longitudinal polarization of vector
meson $J/\Psi$ can be formulated as

\begin{equation}
  \label{eq:dgl}
\frac{{\rm d}\Gamma_L}{{\rm d}q^2}=\frac{G_F^2
\lambda(q^2)^{1/2}|V_{cb}|^2q^2}{192\pi^3 m_{B_c}^3}|H_0(q^2)|^2,
\end{equation}
\begin{equation}
  \label{eq:dgt}
\frac{{\rm d}\Gamma_T}{{\rm d}q^2} =\frac{G_F^2
\lambda(q^2)^{1/2}|V_{cb}|^2q^2}{192\pi^3 m_{B_c}^3}
\left(|H_+(q^2)|^2+|H_-(q^2)|^2\right),
\end{equation}
respectively. Here the helicity amplitudes are expressed as follows:
\begin{equation}
  \label{eq:helamp}
H_\pm(q^2)=\frac{\lambda(q^2)^{1/2}}{m_{B_{c}}+
m_{J/\Psi}}\left[V(q^2)\mp
\frac{(m_{B_{c}}+m_{J/\Psi})^2}{\lambda(q^2)^{1/2}}A_1(q^2)\right],
\end{equation}
\begin{eqnarray}
  \label{eq:h0a}
&& H_0(q^2) =
\frac1{2m_{J/\Psi}\sqrt{q^2}}\left[-\frac{\lambda(q^2)}{m_{B_{c}}
+ m_{J/\Psi}}A_2(q^2)\right.~~~~~\nonumber\\
&&~~\left.+(m_{B_{c}}+m_{J/\Psi})
(m_{B_{c}}^2-m_{J/\Psi}^2-q^2)A_1(q^2)\right] .
\end{eqnarray}
While summing up the various polarizations, the semileptonic
differential decay rate of $B_{c}\to J/\Psi \ell \nu$ over $q^2$ is
obtained
\begin{eqnarray}
  \label{eq:dgv}
&&\frac{{\rm d}\Gamma}{{\rm d}q^2}( B_{c}\to J/\Psi
\ell\nu)=\frac{G_F^2 \lambda(q^2)^{1/2}|V_{cb}|^2q^2}{192\pi^3
m_{B_c}^3} ~~~~~~~~~~~~\nonumber\\
&&~~~~~~~~~\times\left(|H_+(q^2)|^2+|H_-(q^2)|^2
+|H_0(q^2)|^2\right) ,
\end{eqnarray}
with $\lambda(q^2)=(m_{B_c}^2+m_{J/\Psi}^2-q^2)^2-4m_{B_c}^2
m_{J/\Psi}^2$.

Similar as $B_{c}\to \eta_c \ell\nu$, the distribution of NLO
differential decay rate on momentum transfer $q^2$ for $B_{c}\to
J/\Psi \ell\nu$ channel with three sets of different values of $z$
is illustrated in Fig.~\ref{Fig:Semilep}. At the maximum recoil
point $q^2=0$, we obtain a value of $3.21^{+0.53}_{-0.37}\times
10^{-12}|V_{cb}|^2 ~\mathrm{GeV}^{-1}$ for (\ref{eq:dgv}), which is
larger than the value of $0.6\times 10^{-12}|V_{cb}|^2
~\mathrm{GeV}^{-1}$ obtained in nonrelativistic quark model
\cite{cqm}. Except for the enhancement from the NLO K-factor and the
NLO Schr\"{o}dinger wave functions at the origin, the result in LO
in NRQCD calculation is intrinsically bigger than what obtained in
nonrelativistic quark model.

\section{ Theoretical uncertainty\label{sec-decaywidth}}

With the input parameters given in Table~\ref{tab:Bc-lep1} and take
$m_{B_c}=6.273\mathrm{GeV}$ \cite{LHCbj}, one can readily obtain the
decay widths numerically, which are presented in
Table~\ref{tab:decay-widths}. In our calculation, the value of the
$J/\Psi$ wave function squared at the origin is extracted from the
leptonic decay width at the NLO in $\alpha_s$ \cite{NRQCD,Chao},
i.e.,
\begin{eqnarray}
|\psi(0)|_{J/\Psi}^2=\frac{m^2_{J/\Psi }}{16\pi\alpha^2e_{c}^2}\frac{
\Gamma (J/\Psi\rightarrow e^+ e^-)}{(1-4\alpha_s C_F/\pi)},
\end{eqnarray}
and the experimental value $\Gamma (J/\Psi\rightarrow e^+
e^-)=5.55\pm0.14\pm0.02~ \mathrm{keV}$ is used. Note that according
to the heavy quark spin symmetry, at leading order in the typical
velocity $v$ expansion in NRQCD, we have $|\psi(0)|_{\eta_c} =
|\psi(0)|_{J/\Psi}$.

It is found that the main uncertainties of the concerned processes
come from two sources, the heavy quark masses and the
renormalization scale. In the evaluation, we vary the charm quark
mass $m_c=1.4~\mathrm{GeV}$ by $\pm 0.1~\mathrm{GeV}$, the bottom
quark mass $m_b=4.9~\mathrm{GeV}$ by $\pm 0.1~\mathrm{GeV}$ and the
renormalization scale $\mu=4.8~\mathrm{GeV}$ by
$^{+1.2}_{-1.8}~\mathrm{GeV}$. The numerical value of pole mass may
vary in a reasonable range, so we need also to consider the
uncertainty coming from the pole mass. Notice that the pole mass
effect tends to be small in the maximum recoil region, as it should
be.

In Table~\ref{tab:decay-widths}, the decay widths calculated through
other approaches, such as QCD Light-Cone Sum Rules, Quark Model,
Bethe-Salpeter equation and  potential model, are also given. In
comparison with QCD LCSR results, our results are almost treble of
theirs. This is understandable considering the large QCD correction
K factor and the NLO charmonium wave function employed.

To see more clearly the uncertainty remaining in the NLO evaluation,
we calculate the decay width in various momentum transfer squared
region. For light leptons($\ell =e,\mu$), we divide $q^2$ into five
bins in maximum recoil region($0\leq q^2\leq 5~\mathrm{GeV^2}$) and
calculate the semileptonic decay rates separately. The results are
presented in Table \ref{tab:bins}. We find that at small
$q^2$($0\leq q^2\leq 1~\mathrm{GeV^2}$), the longitudinally
polarized $J/\Psi$ events dominate over the transversally polarized
ones by a factor 8.5, and the difference reduces with the $q^2$
increase. While for lepton $\tau$, we divide $q^2$ into two
bins($m^2_\tau\leq q^2\leq 4,~4\leq q^2\leq 5~\mathrm{GeV^2}$) in
maximum recoil region. Here the physical mass of lepton $\tau$ is
taken to be $m_\tau=1.776~\mathrm{GeV}$ \cite{PDG:2010}, and the
results are shown in Table~\ref{tab:bins-tau}.

\section{Conclusions\label{sec-con}}

The NLO semileptonic differential decay rates of $B_c$ meson to
charmomia are analyzed in detail with various choices of parameters.
The uncertainties of patrial decay widths in different bins of
momentum transfer $q^2$ are evaluated. For $B_{c}\to J/\Psi \ell\nu$
process, the partial decay widths for transverse and longitudinal
polarizations are investigated separately. The distribution in the
maximum recoil is found testable in the LHCb experiment, and in turn
the NRQCD factorization will be also testified. Based on certain
model, phenomenologically the form factors are extrapolated to the
minimal recoil region, and we estimate the total rates of $B_c$
semileptonic decay to charmonium.

\hspace{2cm}

{\bf Acknowledgements}:

This work was supported in part by the National Natural Science
Foundation of China(NSFC) under the grants 10935012, 10821063 and
11175249.

\hspace{2cm}

\appendix

\section{The NLO $B_c$ to Charmonia transition form factors}
In this appendix, the QCD NLO $B_c$ to charmonium transition form
factors are given at the first order in power of $m_c/m_b$. For
compactness, we define $z=m_c/m_b$, $s=\frac{m_b^2}{m_b^2-q^2}$, and
$\gamma=\frac{m_b^2-q^2}{4m_bm_c}$. Besides, the form factors at
maximum recoil point, i.e. $q^2=0$, are also presented, which are in
agreement with what given in references \cite{Bell:2007, Sun:2011}.

\begin{widetext}

\begin{table}[t]
\caption{\label{tab:decay-widths} The branching ratios (in $\%$) of
exclusive semileptonic decays of  $B_c$ meson to ground state
charmonia, in comparison with the results of Light-Cone Sum Rules
\cite{Huang:2008,exBc}, Quark Model
\cite{cqm,Lu:2009,narod,RCQM,Ebert:2003}, the calculation of
Bethe-Salpeter equation \cite{Zhang:1994}, and QCD relativistic
potential model \cite{CdF,AbdEl-Hady:1999xh}. In the evaluation, the
$B_c$ lifetime $\tau(B_c) =0.453$ ps, $\ell$ stands $e$ or $\mu$,
$m_c/m_b=1.4/4.9$, $\mu=4.8~\mathrm{GeV}$, and $|V_{cb}|=0.0406$.
The uncertainties in our calculation come from varying the value of
$m_c/m_b$ from $1.5/4.8$ to $1.3/5.0$, varying the renormalization
scale $\mu$ from 3 to $6~\mathrm{GeV}$, and varying the pole mass
$m_{\mathrm{pole}}$ from $4.25$ to $4.75$ $\mathrm{GeV}^2$
\cite{Kiselev:2000,exBc}, respectively.}
\def\arraystretch{1.5}
\begin{center}
\begin{tabular}{|c|c|c|c|c|c|c|c|c|c|c|c|}
\hline
 Mode & This paper  & \cite{Kiselev:2000,exBc}
 &\cite{Huang:2008} & \cite{cqm}& \cite{Lu:2009}&
 \cite{narod}  & \cite{RCQM} &\cite{Ebert:2003}
 &  \cite{Zhang:1994} & \cite{CdF} & \cite{AbdEl-Hady:1999xh}\\
\hline $B_c^- \to \eta_c \ell \nu$     &
$2.1^{+0.5+0.4+0.2}_{-0.3-0.1-0.1}$ & 0.75 &1.64 & 0.48& 0.67 & 0.59
& 0.81& 0.40 &
 0.97  & 0.15 & 0.76   \\
$B_c^- \to \eta_c \tau \nu$  &
$0.64^{+0.07+0.14+0.10}_{-0.08-0.06-0.05}$
& 0.23 &0.49&0.16& 0.19 & 0.20 & 0.22& -     & -& -& -\\
\hline $B_c^- \to J/\psi \ell \nu $ &
$6.7^{+2.1+1.0+0.9}_{-1.2-0.4-0.6}$ & 1.9&2.37 &1.5 &1.49 & 1.20  &
2.07& 1.21 & 2.35&
1.47 & 2.01  \\
$B_c^- \to J/\psi \tau \nu $ &
$0.52^{+0.16+0.08+0.08}_{-0.09-0.03-0.05}$ & 0.48&0.65 &0.4 & 0.37 &
0.34 & 0.49 & - &
-&- &- \\
\hline
\end{tabular}
\end{center}
\end{table}

\begin{table}
\begin{center}
\caption{\label{tab:bins} The NLO partial decay widths for various
$q^2$. For $J/\Psi$, the partial decay widths for
transverse($\varepsilon^{*}_{\perp}$) and longitudinal
($\varepsilon^{*}_{\parallel}$) polarizations are presented
separately.} \vskip 0.6cm
\begin{tabular}{|c|c|c|c|c|c|}
  \hline
bins of $q^2~(\mathrm{GeV^2})$& $0\leq q^2\leq1$ & $1\leq q^2\leq2$
& $2\leq q^2\leq3$ & $3\leq q^2\leq4$ & $4\leq q^2\leq5$
\\\hline
  $\Delta \Gamma(B_{c}\to\eta_c
\ell\nu)~(10^{-15}~\mathrm{GeV})$ & $8.06^{+1.17+1.96}_{-0.77-0.74}$
& $9.73^{+1.78+2.37}_{-1.16-0.89}$  &
$12.0^{+2.78+2.95}_{-1.77-1.11}$ &$15.2^{+4.47+2.73}_{-3.77-1.41}$
&$20.0^{+7.57+5.06}_{-4.36-1.87}$\\\hline
  $\Delta \Gamma(B_{c}\to J/\Psi(\varepsilon^{*}_{\perp})
\ell\nu)~(10^{-15}~\mathrm{GeV})$
&$0.70^{+0.215+0.159}_{-0.141-0.061}$ &
$2.64^{+0.95+0.60}_{-0.60-0.23}$ & $5.84^{+2.54+1.34}_{-1.51-0.51}$&
$11.28^{+6.03+2.61}_{-3.35-1.00}$&
$20.97^{+14.14+4.90}_{-7.16-1.87}$\\\hline $\Delta \Gamma(B_{c}\to
J/\Psi(\varepsilon^{*}_{\parallel})
\ell\nu)~(10^{-15}~\mathrm{GeV})$ & $6.01^{+1.14+1.40}_{-0.78-0.53}$
& $7.87^{+1.93+1.84}_{-1.27-0.70}$&
$10.64^{+3.38+2.51}_{-2.72-0.95}$&$14.96^{+6.18+3.56}_{-3.62-1.35}$
& $22.07^{+12.06+5.31}_{-6.45-2.01}$
\\\hline
  $\Delta \Gamma(B_{c}\to J/\Psi \ell\nu)~(10^{-15}~\mathrm{GeV})$ &
  $6.71^{+1.35+1.56}_{-0.92-0.59}$ & $10.52^{+2.89+2.45}_{-1.88-0.93}$ &
  $16.49^{+5.92+3.86}_{-4.24-1.47}$ &$26.24^{+12.21+6.18}_{-6.98-2.35}$ &
  $43.04^{+26.20+10.22}_{-13.61-3.88}$ \\
  \hline
\end{tabular}
\end{center}
\end{table}

\begin{table}
\begin{center}
\caption{\label{tab:bins-tau} The NLO partial decay widths of
processes $B_{c}\to\eta_c \tau\nu_\tau$ and $B_{c}\to J/\Psi
\tau\nu_\tau$ for various $q^2$, where the maximum recoil point is
at $m^2_\tau$. } \vskip 0.6cm
\begin{tabular}{|c|c|c|}
  \hline
bins of $q^2~(\mathrm{GeV^2})$& $m^2_\tau\leq q^2\leq4$ & $4\leq
q^2\leq5$
\\\hline
  $\Delta \Gamma(B_{c}\to\eta_c
\tau\nu_\tau)~(10^{-15}~\mathrm{GeV})$ &
$2.460^{+0.9245+0.655}_{-0.538-0.241}$ &
$17.62^{+8.42+4.70}_{-4.56-1.73}$  \\\hline
  $\Delta \Gamma(B_{c}\to J/\Psi
  \tau\nu_\tau)~(10^{-15}~\mathrm{GeV})$
  &$0.821^{+0.375+0.194}_{-0.213-0.073}$  &
  $6.922^{+4.017+1.648}_{-2.107-0.625}$ \\
  \hline
\end{tabular}
\end{center}
\end{table}
\begin{eqnarray}
\frac{f^{NLO}_+(q^2)}
{f^{LO}_+(q^2)}&=&1+\frac{\alpha_s}{4\pi}\{\frac{1}{3}
(11 C_A-2 \text{n}_f) \log
 (\frac{\mu^2}{2 \gamma\text{m}_c^2})-\frac{10\text{n}_f}{9}+\frac{(\pi
^2-6 \log (2)) (s-1)+3 s \log (\gamma )}{6 s+3}\nonumber\\
&&+\frac{C_A}{72 s^2-18}\Big(18 s^2 (2 s-1) \log ^2(s)+18 (8 \log
(2)
s^3-2 \log (2) s^2-5 \log (2) s+s\nonumber\\
&&+2 \log (2)) \log (s)+(2 s-1) (268 s+\pi ^2 (6 s^2-3
   s-6)+170)-9 (2 s\nonumber\\
&&-1) \log (\gamma ) (\log (\gamma ) s-(2+2 \log (2)) s+4 \log
(2))+18 (2 s-1) (4 s^2+s\nonumber\\
&&-2) \text{Li}_2(1-2 s)-18 (4 s^3-5
   s+2) \text{Li}_2(1-s)+18 (s (4 s (s+1)-11)\nonumber\\
&&+4) \log ^2(2)-36 (5 (s-1) s+1) \log
   (2)\Big)\nonumber\\
&&+\frac{C_F}{6 (1-2 s)^2 (2 s+1)}\Big(-6 (2 (s-1) s-1) \log ^2(s)
(1-2
s)^2+3 \log (\gamma ) (23 s\nonumber\\
&&+(5 s-2) \log (\gamma )-4 (s+1) \log (2)+12) (1-2 s)^2-12 (4
s^2+s\nonumber\\
&&-2) \text{Li}_2(1-2 s)
   (1-2 s)^2+12 (s (2 s+3)-1) \text{Li}_2(1-s) (1-2 s)^2\nonumber\\
&&-(\pi -2 \pi  s)^2 (s (4 s-19)+4)+3 (-32 \log ^2(2) s^4-4 (69+2
\log (2) (-37\nonumber\\
&&+5 \log (2))) s^3+8 (18+\log
   (2) (-31+9\log (2))) s^2+(61+28 \log (2)\nonumber\\
&&-26 \log ^2(2)) s+12\log (2)+2 \log ^2(2)-32)+(6 s (8 s (s (-4
\log (2) s\nonumber\\
&&+3 \log (2)+3)+2 \log (2)-3)-18
   \log (2)+7)+24 \log (2)) \log (s)\Big)\}\,,
\end{eqnarray}
\begin{eqnarray}
\frac{f^{NLO}_+(0)}{f^{LO}_+(0)}& =
&1+\frac{\alpha_s}{4\pi}\{\frac{1}{3} (11 C_A-2 \text{n}_f) \log
 \left(\frac{2\mu^2}{m_bm_c}\right) -
 \frac{10\text{n}_f}{9}-\frac{1}{3} \log
 \left(z\right)-\frac{2 \log (2)}{3}\nonumber\\
&&+C_F\left(\frac{1}{2} \log ^2\left(z\right)+\frac{10}{3} \log (2)
\log \left(z\right)-\frac{35}{6} \log \left(z\right)+\frac{2 \log
   ^2(2)}{3}\right.\nonumber\\
&&\left.+3 \log (2)+\frac{7 \pi ^2}{9}-\frac{103}{6}\right)\nonumber\\
&&+C_A\left(-\frac{1}{6} \log ^2\left(z\right)-\frac{1}{3} \log (2)
\log \left(z\right)-\frac{1}{3} \log \left(z\right)+\frac{\log
   ^2(2)}{3}\right.\nonumber\\
&&\left.-\frac{4 \log (2)}{3}-\frac{5 \pi
^2}{36}+\frac{73}{9}\right)\}\,,
\end{eqnarray}
\begin{eqnarray}
\frac{f^{NLO}_0(q^2)}{f^{LO}_0(q^2)}& =
&1+\frac{\alpha_s}{4\pi}\{\frac{1}{3} (11 C_A-2 \text{n}_f)
\log(\frac{\mu^2}{2 \gamma\text{m}_c^2})
-\frac{10\text{n}_f}{9}+\frac{\log (\gamma )}{3}\nonumber\\
&&+\frac{C_A}{36 s-18}\Big(-6 \text{Li}_2(1-s) (1-2 s)^2+6 \log
^2(2) (1-2 s)^2+6 s (2 s-1) \log ^2(s) \nonumber\\
&&+(2 s-1) \left(\pi ^2 (2 s-3)+146\right)+(12 s \log (2) (4 s-3)+6
\log (2)-6) \log (s) \nonumber\\
&&-3 (2 s-1) \log (\gamma ) (\log (4 \gamma )-2)+6 \left(8 s^2-6
s+1\right) \text{Li}_2(1-2 s) -12 s \log (2)\Big)
\nonumber\\
&&+\frac{C_F}{18 (1-2 s)^2 (s-1)}\Big(-6 (s-1) (2 s-3) \log ^2(s)
(1-2
s)^2\nonumber\\
&&-12 (s-1) (4 s-1) \text{Li}_2(1-2 s) (1-2 s)^2 +24
\left(s^2-1\right) \text{Li}_2(1-s) (1-2 s)^2\nonumber\\
&&-6 (-6 s (2 s (3 s-8)+11)+2 s (4 s (s (4 s-9)+7)-9) \log (2) +2
\log (2)+13) \log (s)\nonumber\\
&&+(s-1) (3 \log (\gamma ) (3 \log (\gamma )-8 \log (2)+35) (1-2
s)^2 -24 (s+2) \log ^2(2) (1-2 s)^2\nonumber\\
&&-(2 s-1) (546 s+\pi ^2 \left(8 s^2-34 s+15\right) -279)+24 (s (43
s-42)+10) \log (2))\Big)\}\,,
\end{eqnarray}
\begin{eqnarray}
\frac{f^{NLO}_0(0)}{f^{LO}_0(0)}&=&\frac{f^{NLO}_+(0)}{f^{LO}_+(0)}\,,
\end{eqnarray}
\begin{eqnarray}
\frac{V^{NLO}(q^2)}{V^{LO}(q^2)}&=&1+\frac{\alpha_s}{4\pi}\{\frac{1}{3}
(11 C_A-2 \text{n}_f) \log
 (\frac{\mu^2}{2 \gamma\text{m}_c^2})-\frac{10\text{n}_f}{9}\nonumber\\
&&-\frac{C_A}{36 s-18}\Big(9 s (2 s-1) \log ^2(s)+18 (2 s \log (2)
(2
s-1)+1) \log (s)\nonumber\\
&&+3 \pi ^2 (s+2) (2 s-1)-2 s (-18 \log ^2(2) s+9 \log ^2(2)+45 \log
(2)+134)\nonumber\\
&&+9 (2 s-1)
   (\log (\gamma )-3) \log (\gamma )+18 s (2 s-1) (2
   \text{Li}_2(1-2 s)-\text{Li}_2(1-s))\nonumber\\
&&+63 \log
   (2)+134\Big)\nonumber\\
&&+\frac{C_F}{6 (1-2 s)^2 (s-1)}\Big(6 (s^2-1) \log ^2(s) (1-2
s)^2+24
(s-1) s \text{Li}_2(1-2 s) (1-2 s)^2\nonumber\\
&&+3 (2 s (s (4 s (4 \log (2) s-8 \log (2)+3)+20 \log (2)-17)-4 \log
(2)+7)-1)
   \log (s)\nonumber\\
&&+(s-1) (6 \log (\gamma ) (\log (\gamma )-6 \log (2)+5) (1-2 s)^2+6
(2 s-9) \log ^2(2) (1-2 s)^2\nonumber\\
&&+(2 s-1) (-204 s+2 \pi ^2 (2
   s^2+s-1)+105)+6 (s (68 s-67)+16) \log (2))\nonumber\\
&&-12 (2 s^2-3 s+1)^2
   \text{Li}_2(1-s)\Big)\}\,,
\end{eqnarray}
\begin{eqnarray}
\frac{V^{NLO}(0)}{V^{LO}(0)}&=&1+\frac{\alpha_s}{4\pi}\{\frac{1}{3}
(11 C_A-2 \text{n}_f) \log
 \left(\frac{2\mu^2}{m_bm_c}\right)-\frac{10\text{n}_f}{9}\nonumber\\
&&+C_F\Big(\log ^2\left(z\right)+10 \log (2) \log \left(
z\right)-5 \log \left(z\right)+9 \log ^2(2)\nonumber\\
&&+
7 \log (2)+\frac{\pi ^2}{3}-15\Big)\nonumber\\
&&+C_A\Big(-\frac{1}{2} \log ^2\left(z\right)-2 \log (2) \log
\left(z\right)-\frac{3}{2} \log
\left(z\right)-3 \log ^2(2)\nonumber\\
&&-\frac{3 \log
   (2)}{2}-\frac{\pi ^2}{3}+\frac{67}{9}\Big)\}\,,
\end{eqnarray}
\begin{eqnarray}
\frac{A_1^{NLO}(q^2)}{A_1^{LO}(q^2)}
=\frac{A_2^{NLO}(q^2)}{A_2^{LO}(q^2)}=
\frac{V^{NLO}(q^2)}{V^{LO}(q^2)}\,,
\end{eqnarray}
\begin{eqnarray}
\frac{A_0^{NLO}(q^2)}{A_0^{LO}(q^2)}&=&1+\frac{\alpha_s}{4\pi}\{\frac{1}{3}
(11 C_A-2 \text{n}_f) \log
 (\frac{\mu^2}{2 \gamma\text{m}_c^2})-\frac{10\text{n}_f}{9}\nonumber\\
&&+\frac{C_A}{72(s-1)s(2 s-1)}\Big(-9 (s-1) s (2 s-1) (2 s+1) \log
^2(s)-9 (2 s (2 s (\log (2) (4
s^2-9)\nonumber\\
&&+3)+12 \log (2)-3)-4 \log (2)-2) \log (s)+(s-1) (-18 (2 s-1) \log
^2(2) (s (2 s+9)-4)\nonumber\\
&&-(2 s-1) (s (3 \pi ^2 (2 s+9)-608)+36)-9 (2 s-1)
\log (\gamma ) (2 \log (\gamma ) s+8 \log (2) s-6 s\nonumber\\
&&+\log (\gamma )-4 \log (2)-3)+9 (4 s (13 s-7)-3)
\log (2))-36 (s (4 s^3-9 s+6)-1) \text{Li}_2(1-2 s)\nonumber\\
&&+18(s-1) (2 s-1) (s (2 s+5)-2) \text{Li}_2(1-s)\Big)\nonumber\\
&&+\frac{C_F}{24 s (2 s^2-3 s+1)^2}\Big(2 \pi ^2 (1-2 s)^2 (s
(2 s-1)+3) (s-1)^2\nonumber\\
&&+24 (1-2 s)^2 (s (2 s+3)-1) \text{Li}_2(1-2 s) (s-1)^2+6 s (2 s+5)
\left(2 s^2-3 s+1\right)^2 \log ^2(s)\nonumber\\
&&+3 (s (s (2 s \log (2) (2 s (76 s-193)+289)+4 s (-120 s^2+369
s+\left(8 s^3-52 s^2+90 s-43\right) \log ^2(2)\nonumber\\
&&-406)-84 \log ^2(2)-28 \log (2)+747)+92 \log ^2(2)-110 \log
(2)-116)+16 \log ^2(2)\nonumber\\
&&+4 (7-9 \log (2)) \log (2)-3)+3 (s (2 s (s (2 s (4 s
(4 \log (2) s-6 \log (2)+6)-28 \log (2)-69)\nonumber\\
&&+156 \log (2)+113)-2 (13+58 \log (2)))+72 \log
(2)-5)-8 \log (2)+1) \log (s)\nonumber\\
&&-6 \left(2 s^2-3 s+1\right)^2 (16 \log (2) s-22 s-2 \log (\gamma
)+4 \log (2)+1) \log (\gamma )\nonumber\\
&&-12 \left(2 s^2-3
   s+1\right)^2 \left(2 s^2+s-2\right) \text{Li}_2(1-s)\Big)\}\,,
\end{eqnarray}
\begin{eqnarray}
\frac{A_0^{NLO}(0)}{A_0^{LO}(0)}&=&1+\frac{\alpha_s}{4\pi}\{\frac{1}{3}
(11 C_A-2 \text{n}_f) \log
(\frac{2\mu^2}{m_bm_c})-\frac{10\text{n}_f}{9}
+C_F\Big(\frac{1}{2} \log ^2(z)-\frac{119}{8}\nonumber\\
&&+7 \log (2) \log (z)-\frac{21}{4} \log (z)+7 \log
^2(2)+\frac{15 \log (2)}{4}\Big)\nonumber\\
&&+C_A\Big(-\frac{3}{8} \log ^2(z)-\log (2) \log (z)-\frac{9}{8}
\log (z)-\frac{7 \pi ^2}{24}+\frac{67}{9}\nonumber\\
&&-\frac{9 \log ^2(2)}{4}+\frac{3 \log (2)}{8}\Big)\}\,.
\end{eqnarray}
\end{widetext}

\end{document}